\documentclass[twoside]{LCWS11}
\usepackage[latin1]{inputenc}
\usepackage[dvips]{graphicx,epsfig,color}
\usepackage{wrapfig,rotating}
\usepackage{amssymb,amsmath,array}
\usepackage{cite}

\begin{document}
\title{
Air cooling for Vertex Detectors} 
\author{ Arantza Oyanguren 
\vspace{.3cm}\\
IFIC (Universidad de Valencia - CSIC) \\
Edificio Institutos de Investigaci\'on. Apartado de Correos 22085\\
E-46071 Valencia  - Spain \\
E-mail: Arantza.Oyanguren@ific.uv.es\\
}

\maketitle

\begin{abstract}

The vertex detectors are crucial detectors for future linear $e^+e^-$
colliders since they must give the most accurate location of any outgoing
charged particles originating from the interaction point. The DEPFET
collaboration is developing a new type of pixel sensors which
provide very low noise and high spatial resolution. In order to precisely
determine the track and vertex positions, multiple scattering in the
detector has to be reduced by minimizing the material in the sensors, 
cooling, and support structures. A new method of cooling by blowing air over
the sensors has been developed and tested. It is applied in the
design and construction of the Belle-II detector and may be used in the
new generation of vertex detectors for linear colliders.

\end{abstract}

\section{Introduction}

High-performance vertex detectors are critical for experiments at future linear colliders, 
since the physics reach at these colliders will depend on the capability to identify and 
distinguish among heavy quarks in multi-jets events, and on the high resolution for tracking 
and vertex reconstruction in very dense jets. 
For this purpose vertex detectors are required to have high point resolution ($<5\mu$m), 
very low material ($X_0\sim 0.3 \%$), high radiation tolerance 
($\sim$ kRad/year), fast integration time ($25-100\mu$s), low occupancy ($<1\%$) and 
low power consumption ($\sim$mW/cm$^2$). 
One of the technologies able to match these requirements is the DEPleted Field Effect Transistor 
(DEPFET), an active pixel sensor with p-channel FETs integrated in a fully depleted 
bulk \cite{depfet}. The charge of a particle traversing the sensor is collected 
in an internal gate created by an n-inplant beneath the transistor channel, 
leading to a modulation of the channel current of the 
transistor depending on the amount of collected charge. 
These devices thus provide in-pixel 
amplification, combined with low noise due to the small capacitance of the internal gate. 
The amplification combined with the low noise leads to excellent signal to noise ratios, 
allowing the construction of thinned detectors to reduce the material budget. 
The power consumption of the active area of DEPFET sensors is very low since the pixels 
passively collect charge and only need power during the readout cycle. 
The pixels are read out row-wise in a rolling shutter mode and are cleared after readout. 
One of the main issues in order to achieve a low material budget and due power consumption 
is to have the proper mechanical design, support, and cooling of the detector. 

\section{The DEPFET pixel detector at Belle-II (PXD)}

Due to their good performance and maturity, DEPFET devices are excellent candidates for the ILC 
vertex detector. In fact they have been the chosen technology for the pixel vertex detector 
of the Belle-II experiment \cite{tdrbelleII}, presently under construction. Requirements for 
a vertex detector at Belle-II and at a Linear Collider are similar although differ in some 
aspects since typical particle momenta, background, and collision rates are different in both 
experiments. 
At Belle-II the quasi-continuous beam with 2 ns between bunch crossings and 100$\%$ duty 
cycle combined with high background rates requires fast continuous readout. 
At the ILC, the two orders of magnitude larger bunch to bunch spacing and the bunch train 
structure of the beam, place less stringent requirements on the readout speed of the 
detector and on the power consumption. 
Pixel size and number are also different at Belle-II and ILC since multiple scattering 
limits the point resolution, leading larger and less pixels per module in the case of 
Belle-II. 
At ILC, a power pulsing of the electronics is envisaged while continuous running is required at Belle-II. 
In both cases, cooling by flowing cold gas over the active area in addition to an active cooling 
is needed to reduce the material budget. 
In Table \ref{table:ilc} the main required features for both experiments are shown. 

\begin{wraptable}{l}{0.67\columnwidth}
\footnotesize
\centerline{\begin{tabular}{|l|l|l|}
\hline
 & Belle-II & ILC \\
\hline
\hline
Point resolution & 10$\mu$m & 5$\mu$m \\
Material budget & $\sim 0.1\% X_0$ &  $\sim 0.1\% X_0$\\
Radiation Tolerance & $>$1MRad/year & $<$100kRad/year \\
Frame time & 10$\mu$s & 25-100$\mu$s \\
Occupancy & 0.4 hits/$\mu\rm{m}^2/s$ &  0.13 hits/$\mu\rm{m}^2/s$ \\
Power consumption & 18W/ladder & 5W entire detector \\
 & (360W entire detector) & (duty cycle 1:200) \\
\hline
\end{tabular}}
\caption{ {\small Features for DEPFET devices at Belle-II and ILC vertex detectors.
 The low power consumption of the ILC detector is expected from the 1:200 duty cycle of the ILC engine.}}
\label{table:ilc}
\end{wraptable}

\begin{figure}
\includegraphics[width=0.5\columnwidth]{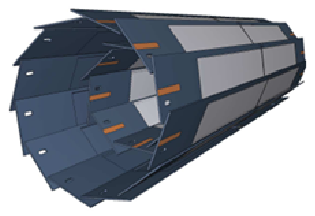}
\includegraphics[width=.4\columnwidth]{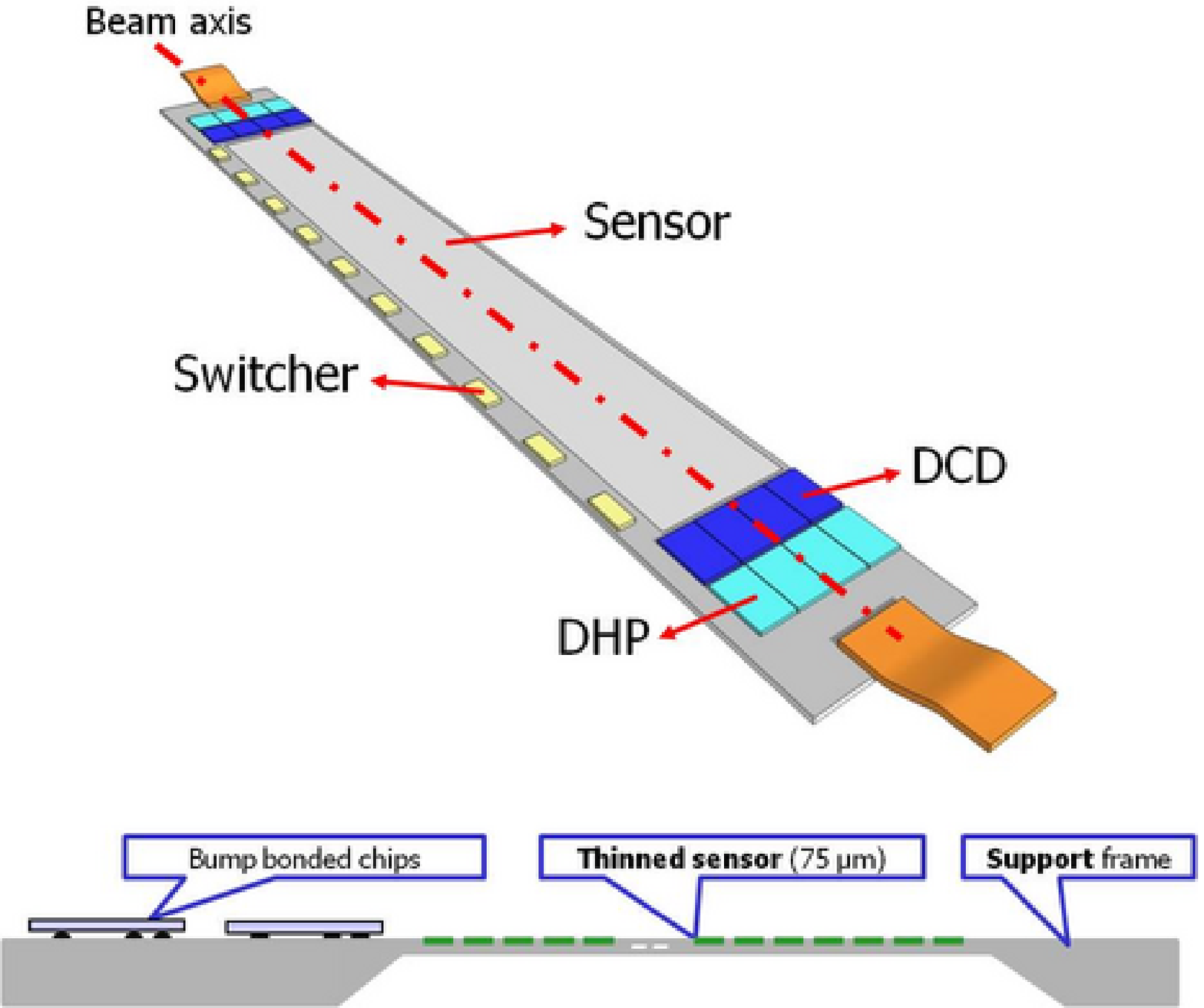}
\caption{Left: Layout of the Belle-II PXD. Right: Schematic view of a DEPFET ladder.\label{fig:ladder}}
\end{figure}

The pixel vertex detector for Belle-II (PXD) consists of two layers at radii 14mm and 22mm, 
with 8 inner and 12 outer ladders, respectively (see Figure \ref{fig:ladder} left)\footnote{For 
ILC 5 layers with 10,11,12,16,20 ladders at radii between 15mm and 60mm are proposed.}.

The central part of the ladders, the active area, is thinned down to 75$\mu$m to reduce the 
material budget. A 450$\mu$m rim gives mechanical support and is where the front-end electronics
is bump bonded. The row control is provided by the steering chips {\it switchers} \cite{fee}, located 
in a lateral balcony, while analog front-end and digital ASICS processors are on the end 
of the stave, on both sides of the ladder ({\it DHP} and {\it DCD} chips \cite{fee,dcd}) . 
Figure \ref{fig:ladder} (right) shows the schematic view of a DEPFET ladder for Belle-II.   

\section{DEPFET PXD Mock-up}

\begin{wrapfigure}{r}{0.4\columnwidth}
\centerline{\includegraphics[width=.4\columnwidth]{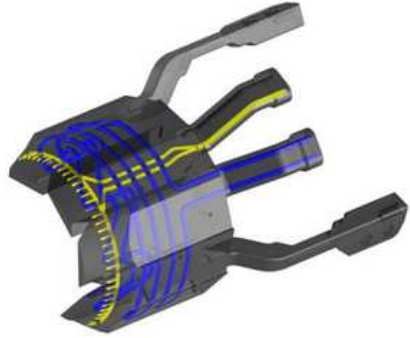}}
\caption{Cooling block structure.}\label{fig:support}
\end{wrapfigure}

One important issue at these devices is the power dissipation and the proper cooling along the ladders.
In the Belle-II configuration, the power dissipation per ladder is about 8W at each extreme due to the {\it DHP} and {\it DCD} modules, 
1W due to the {\it switchers}, and about 1W in the sensor region. The ladders are mechanically supported in the edges by 
stainless steal cooling block structures attached to the beam pipe on each side (see Figure \ref{fig:support}). 
These structures have integrated conduits for CO${_2}$ circulation (blue channels in Figure \ref{fig:support}) 
to dissipate the power of the {\it DCD} and {\it DHP} chips at the end of the staves, and similar embedded pipes with a 
cold air distribution ring (yellow channels in Figure \ref{fig:support}) which have to provide 
cooling by convection between the inner and outer ladders, to dissipate the heating generated by 
the sensor and {\it switcher} modules. The air cooling aims in addition to avoid large temperature gradients along 
the ladders which could cause local strains being dangerous for the sensors.        

\begin{figure}
\centerline{\includegraphics[width=.5\columnwidth]{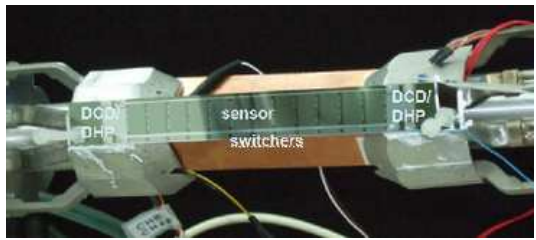}}
\caption{Thermo-mechanical mock-up}\label{fig:mockup}
\end{figure}

With the purpose of testing the mechanical design and cooling for the PXD detector at Belle-II, and in particular 
the air cooling mechanism for the DEPFET sensors, a thermo-mechanical mock-up has been set up. It consist of an 
aluminum beam pipe and four support and cooling block structures. The cooling blocks are 
fabricated by a 3D sintering process and several materials have been evaluated \footnote{Among CL 20ES stainless steal, 
EOS-AlSi10Mg, EOS-NiBr-DM20, EOS-stainless steal-GP1,  EOS-CoCr-MP1, the best solution seems to be CL 20ES, taking into account thermal 
and mechanical properties.}. The cooling blocks are cooled down with CO$_2$ at 12bar, using a CO$_2$ open system,
reaching temperatures of about $ -30^\circ$C. For convective cooling, dry air and N$_2$ gas flows are used, cooled down in 
the atmosphere of a liquid $N_2$ dewar, reaching air flow temperatures around -10$^\circ$C. Several dummy ladders are used in the tests 
(Cu and Al ladders with heaters and resistive silicon samples). The mock-up is kept inside a thermally isolated box with dry atmosphere
(humidity level below 10$\%$). Figure \ref{fig:mockup} shows the cooling blocks with copper inner ladders and a Si resistive outer sample.       

\section{Air cooling studies}

Temperature profiles are measured along inner and outer ladders and in the cooling blocks with an IR camera (ThermaCAM$^{{\rm TM}}$ SC500, FLIR Systems) and PT100 probes. 
Images from the IR camera have to be corrected for material emissivities. 
An emissivity calibration for several materials has been carried out, 
using temperatures between -20$^\circ$C and 90$^\circ$C. In Figure \ref{fig:ladder_on} a thermal image acquired when the resistive ladder is switched on can be seen\footnote{Note that emissivities $\epsilon$ are different for different materials and in the image only materials with $\epsilon \sim 1$ correspond to the temperature scale.}.

The effect of the CO$_2$ and air convection cooling has been studied: 
Analysis of the temperatures reached by the cooling blocks and the power dissipation of the {\it DCD} and {\it DHP} chips once the cooling 
structures are cooled down with CO$_2$ are performed.  
For the air convection cooling, the effect of the air flow at several temperatures on the power dissipation for the sensor and {\it switchers} 
has been evaluated.     

\section{Results}
\begin{wrapfigure}{r}{0.56\columnwidth}
\centerline{\includegraphics[width=.5\columnwidth]{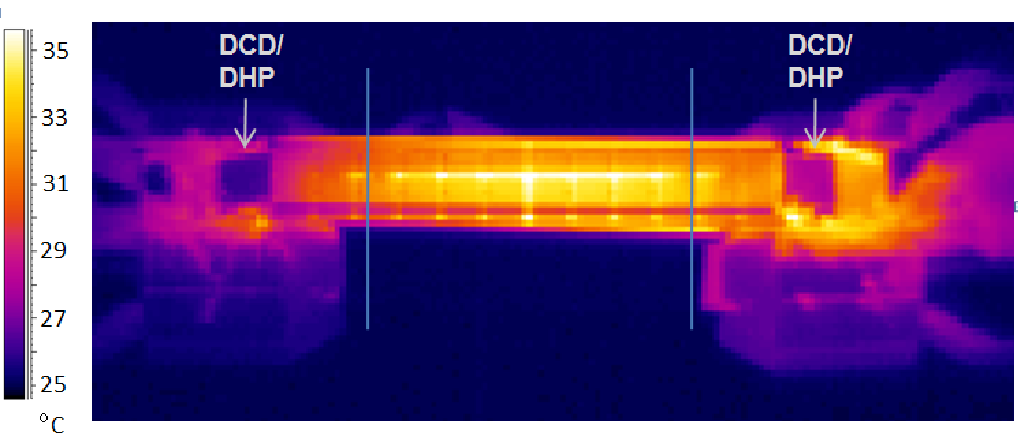}}
\caption{Thermal image with the resistive sample switched on. The power dissipation is about 2W for the sensor, 0.25W for the {\it switchers} and 2.5W at each edge of the ladder.}\label{fig:ladder_on} 
\centerline{\includegraphics[width=.5\columnwidth]{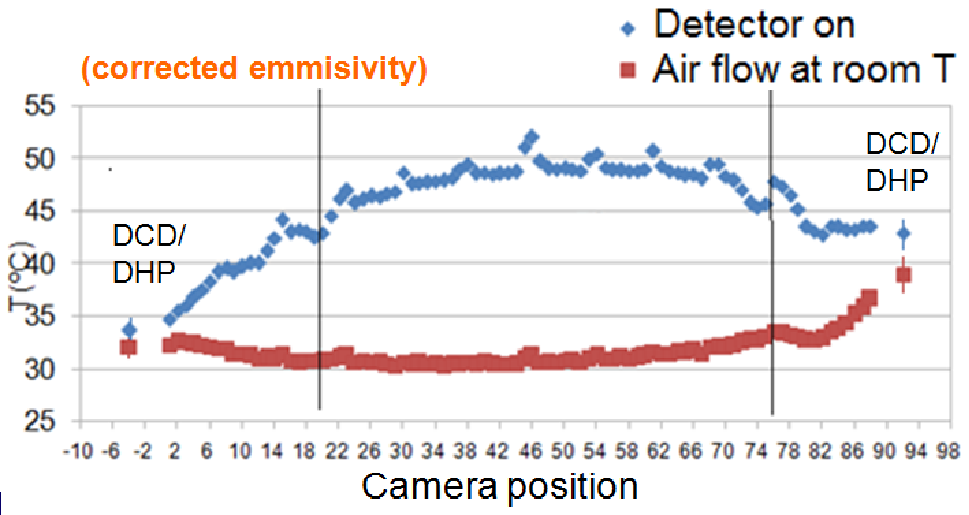}}
\caption{Results of blowing air at room temperature. The cooling blocks are not cooled down.}\label{fig:graph0}
\end{wrapfigure}

Figure \ref{fig:graph0} shows the importance of the air convection cooling 
for the power dissipation of the ladder. 
Even at room 
temperature ($25^\circ$C), without cooling down the cooling blocks, the air flow 
decreases and homogenizes the temperature along the detector. \\
The temperature decreases about 15$^\circ$C for a 2W power dissipation in the sensor region. The air flow also 
reduces the temperature gradient from 18$^\circ$C to 8$^\circ$C along the detector. 
This is true even for small air flow velocities, as it was argued in reference \cite{carlos}. 

\begin{figure}
\centerline{\includegraphics[width=.5\columnwidth]{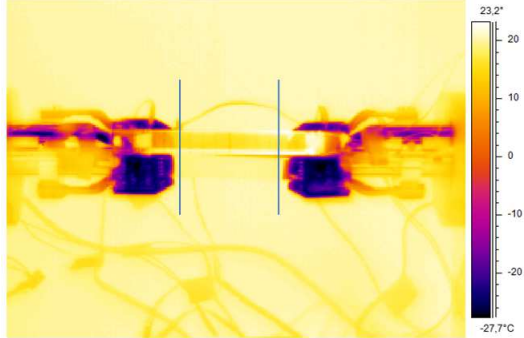}}
\caption{Thermal image when the cooling blocks are cooled down with CO$_2$.}\label{fig:ladder_cold}
\end{figure}

\begin{wrapfigure}{r}{0.52\columnwidth}
\centerline{\includegraphics[width=.5\columnwidth]{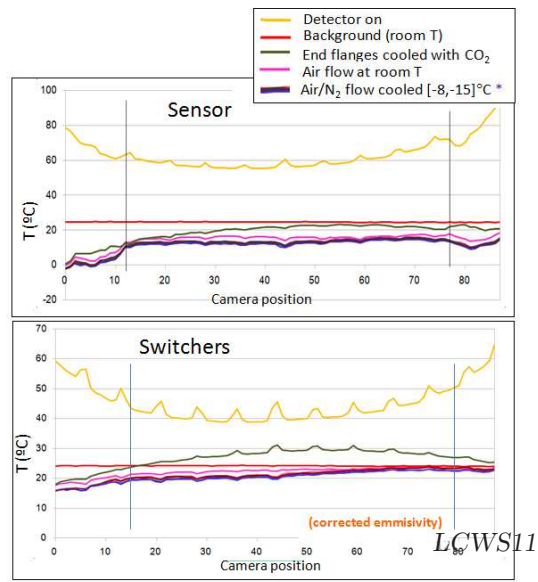}}
\caption{Effect of CO$_2$ and air flow cooling on the sensor and {\it switchers} regions.}\label{fig:graphs}
\end{wrapfigure}

Results when cooling down the cooling blocks with CO$_2$ and blowing dry air and N$_2$ gas at several temperatures are shown in Figures \ref{fig:ladder_cold} and \ref{fig:graphs}.
 The power dissipation at the edges of the ladder has been set to 8W per side, while the sensor and { \it switchers} regions are 1W and 0.5W, respectively. The temperature of the sensor reaches about 60$^\circ$C, decreasing below room temperature with the CO$_2$ cooling. The cold air flow decreases and homogenizes the temperature of the sensor at about 15$^\circ$C, with a $\Delta T_{max}$ 
along the sensor of about $10^\circ$C. In the {\it switcher} region only the right {\it switcher} was operating, but the effect of the cold air flow shows 
similar results. Together with the CO$_2$ cooling, it decreases the {\it switchers} temperature below room temperature with a maximum  
$\Delta T$ along the ladder below $10^\circ$C.  

 Preliminary studies of ladder vibrations coming from the air flows have been performed using displacement capacitive sensors.
The maximum displacements of the ladders is about 4$\mu$m in the central part when the air is blowing, and no large amplitude vibrations have been observed.   

\section{Conclusions}
Convective cooling by blowing cold air between ladders seems to be a promising method to 
cool down and homogenize temperatures of vertex detectors without adding new material sensible 
to multiple scattering, thus improving the tracking resolution.  
This technique is being used for DEPFET sensors which will be installed at the BELLE-II experiment. 
The high-performance DEPFET technology is candidate for the vertex detector at a future Linear Collider. 
The air flow cooling mechanism may be suitable at this and other facilities, improving the power dissipation of these and 
similar devices.     

\begin{footnotesize}


\end{footnotesize}


\end{document}